\documentclass[sigconf]{acmart}
\usepackage{natbib}
\AtBeginDocument{%
  }

\setcopyright{acmlicensed}
\copyrightyear{2025}
\acmYear{2025}
\acmDOI{XXXXXXX.XXXXXXX} 

\begin{document}

\title{Repeton: Structured Bug Repair with ReAct-Guided Patch-and-Test Cycles}
\author{Nguyen Phu Vinh}
\affiliation{%
  \institution{Uppsala University}
  \country{Sweden}
}

\author{Anh Chung Hoang}
\affiliation{%
  \institution{Hanoi University of Science \& Technology}
  \country{Vietnam}
}

\author{Chris Ngo}
\affiliation{%
  \institution{Knovel Engineering Lab}
  \country{Singapore}
}

\author{Truong-Son Hy}
\affiliation{%
  \institution{The University of Alabama at Birmingham}
  \country{USA}
}








\begin{abstract}

Large Language Models (LLMs) have shown strong capabilities in code generation and comprehension, yet their application to complex software engineering tasks often suffers from low precision and limited interpretability. We present \textbf{Repeton}~\footnote{\url{https://github.com/phuvinhnguyen/Repeton}}, a fully open-source framework that leverages LLMs for precise and automated code manipulation in real-world Git repositories. Rather than generating holistic fixes, \textbf{Repeton} operates through a structured patch-and-test pipeline: it iteratively diagnoses issues, proposes code changes, and validates each patch through automated testing. This stepwise process is guided by lightweight heuristics and development tools, avoiding reliance on embedding-based retrieval systems. Evaluated on the SWE-bench Lite benchmark, our method shows good performance compared to RAG-based methods in both patch validity and interpretability. By decomposing software engineering tasks into modular, verifiable stages, \textbf{Repeton} provides a practical path toward scalable and transparent autonomous debugging.

\end{abstract}

\ccsdesc[500]{Software and its engineering~Software creation and management~Software V\&V~Software testing and debugging~Debugging}
\ccsdesc[300]{Computing methodologies~Artificial intelligence~Natural language processing}
\ccsdesc[100]{Software and its engineering~Software organization and properties~Software functional properties~Correctness}
\ccsdesc[100]{Computing methodologies~Artificial intelligence~Knowledge representation and reasoning~Reasoning about belief and knowledge}

\keywords{Automated Software Engineering, LLM Agents, Code Generation, Bug Fixing, Program Repair, Git Automation, Iterative Refinement, Tool-Augmented LLMs}

\received{8 June 2025}

\maketitle

\section{Introduction}

Software debugging remains one of the most labor-intensive and cognitively demanding tasks in software development. As codebases grow in size and complexity, diagnosing and fixing bugs increasingly requires deep domain knowledge, iterative testing, and contextual reasoning. Recent advances in large language models (LLMs) have created new opportunities to automate parts of this process. LLMs have demonstrated strong performance in code generation, understanding, and repair, improving agents' ability in automating debugging workflows. However, many pipelines and frameworks in software debugging are closed-source or rely on closed-source LLMs, which restricts the development of this field and causes problems such as the inability to deploy the pipeline locally.

\begin{figure*}[htpb]
    \centering
    \includegraphics[width=\textwidth]{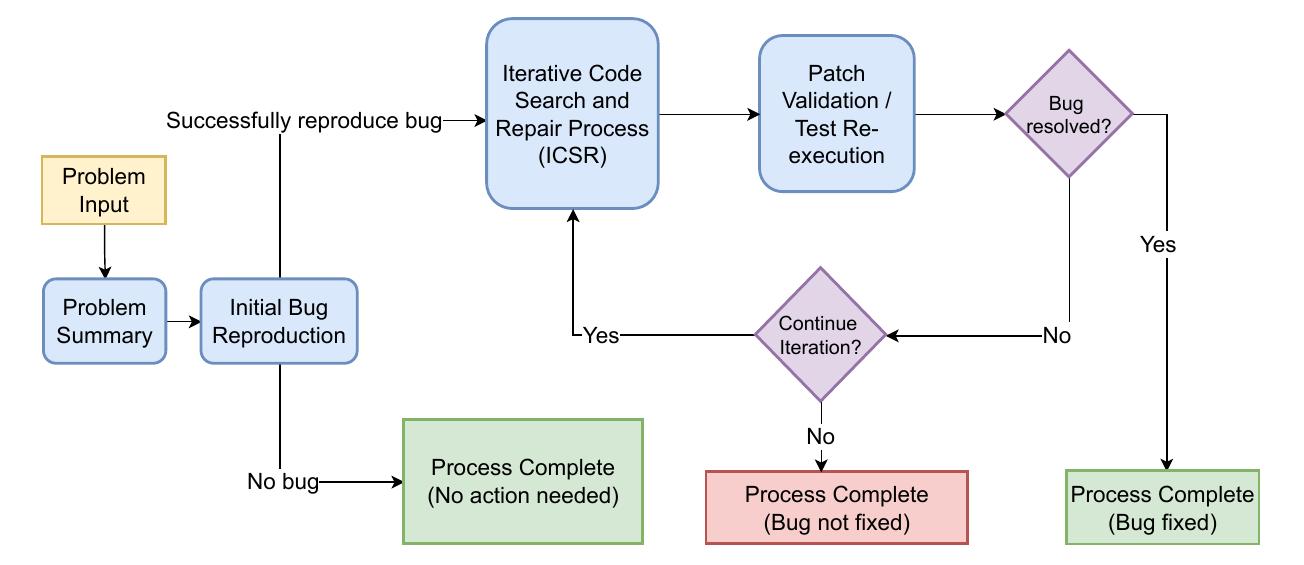}
    \caption{Iterative Repair and Validation workflow. The flow combines code search and patch generation with error reproduction to ensure each patch resolves the bug without introducing new errors.}
    \label{fig:irv-flow}
\end{figure*}

In this work, we introduce a fully open-source debugging agent designed to overcome these limitations through a modular, embedding-free pipeline that uses completely open-source LLMs solely for structured reasoning. Instead of relying on closed-source LLM, our system guides the agent through a step-by-step process inspired by how developers debug: understanding and summarizing the problem, reproducing errors, narrowing the search space with keyword heuristics, identifying suspect files and functions, and iteratively generating and testing patches. This pipeline, illustrated in Figure~\ref{fig:irv-flow}, organizes the agent’s reasoning in a transparent and interpretable way, combining static inspection with runtime feedback.

Our approach focuses on practicality and efficiency. It avoids neural embeddings and external retrieval systems, such as some existing open-weight methods, instead using symbolic techniques such as keyword extraction, structured file search, and test-driven validation. This reduces the complexity of the framework and also reduces the time to embed the entire codebase while still being flexible enough to handle diverse software projects. After that, to evaluate our system, we use SWE-bench Lite, a benchmark of real-world GitHub issues that require multi-file reasoning and test-based validation, to assess and compare our method with some existing open-weight solutions. Our results show that the proposed agent performs well compared to solely RAG-based methods despite being unable to surpass SweFixer. Lastly, we analyze failure cases of our framework, which can provide useful information for further research in this field. This work provides a reproducible and transparent foundation for building autonomous debugging agents. All components, including model weights, pipeline code, and evaluation scripts, are open-sourced to support further research and real-world adoption.

\section{Related Works}

Rapid advancement of large language models (LLMs) has led to the emergence of coding agents, systems that go beyond simple code generation to engage in autonomous software development tasks. Models like Codex~\cite{chen2021evaluating} and CodeT5+~\cite{wang2023codet5plus} demonstrate strong performance on code completion and synthesis benchmarks, while frameworks such as GPT-Engineer~\cite{osika2023gptengineer} and Auto-GPT~\cite{yang2023autogpt} showcase the potential of chaining LLM actions to build software from natural language specifications. However, these systems typically target isolated or synthetic tasks and struggle to handle the complexity of real-world software projects that require multistep reasoning, context tracking, tool usage, and iterative testing.

This gap has raised interest in building autonomous agents that can perform end-to-end bug fixing across entire codebases. Rather than just generating snippets, these agents can identify relevant files, interpret issues, locate faults, and fix the problem. In response to this broader challenge, previous work explores a spectrum of methods and tools. Classical approaches in automated program repair (APR) separate fault localization from patch generation, using techniques such as spectrum-based analysis, mutation testing, and symbolic synthesis (e.g., GenProg~\cite{legoues2012genprog}, Angelix~\cite{mechtaev2016angelix}). Learning-based methods treat repair as translation from buggy code to fixed code, but often fail to generalize beyond training distributions.

More recent systems adopt agentic architectures to coordinate these steps. SWE-agent~\cite{yang2024sweagent} and FixAgent~\cite{lee2024fixagent} organize specialized agents based on LLM (e.g., bug explainer, fault locator, patch proposer) that collaboratively solve debugging tasks. RepairAgent~\cite{bouzenia2024repairagent} and AutoCodeRover~\cite{zhang2024autocoderover} treat the LLM as a high-level planner that invokes tools such as search, compilation, and testing. VulDebugger~\cite{liu2025vuldebugger} combines dynamic run-time information with LLMs to guide security patching. These systems show that agentic design enables more robust debugging, especially when combined with static and dynamic analysis.

To rigorously evaluate such agents, benchmarks such as the SWE benchmark~\cite{jimenez2024swebench} have emerged as critical infrastructure. Derived from real GitHub issues and pull requests, SWE-bench tests agents on realistic bug-fixing tasks involving complex, multi-file projects and test-driven validation. It has become the standard benchmark for assessing autonomous repair agents, with a curated subset (SWE-bench Lite) introduced for efficient experimentation. Performance in the SWE benchmark reveals the limitations of mand LLMs, with top models solving only a fraction of cases, highlighting the challenge of scaling debugging to real-world codebases.

Given these challenges, open-source software agents have become increasingly important. SWE-Fixer~\cite{xie2025swefixer}, built on InternLM, demonstrates competitive performance using structured prompting, memory modules, and tool-augmented reasoning without relying on closed APIs. Moatless Tools~\cite{aorwall2024moatlesstools} emphasizes simplicity and reproducibility, offering lightweight, interpretable primitives for debugging without embedding-heavy infrastructure. Our approach shares this open, modular philosophy that extends it with runtime feedback loops and symbolic heuristics to minimize inference overhead. The SWE-agent project~\cite{yang2024sweagent} further exemplifies open innovation, allowing LLMs to autonomously fix GitHub issues with pluggable model backends, achieving strong results with models like LLaMA-32B. In contrast, proprietary systems like Cognition Labs' Devin~\cite{wu2024devin} remain closed-source, despite their reported performance (13.9\% on SWE-bench). As such, open-source debugging agents not only promote transparency and reproducibility but are essential for democratizing progress in AI-assisted software engineering. Our work builds on this ethos, advancing fully open agents for practical, scalable, and interpretable automated debugging.

\section{Methods}
\label{sec:method}

This section introduces \textbf{Repeton}, a framework powered by the Large Language Model (LLM) designed for the rectification of autonomous and precise codes.

\subsection{Iterative Repair and Validation (IRV)}
\label{ssec:irv_loop}

The Iterative Repair and Validation (IRV) process represents the core \textit{patch-and-test} strategy of \textbf{Repeton}, orchestrated through the \texttt{Testing} and \texttt{Patching} modules. The process begins by summarizing the reported issue using a large language model (LLM), which distills the original problem description into a concise summary denoted \(P_{\text{sum}}\). This summary serves as a guiding context for both patch generation and testing.

The \texttt{Testing} module uses a ReAct-style~\cite{yao2023reactsynergizingreasoningacting} reasoning flow to iteratively build a test program that can reproduce the reported bug. Using both the original problem description and its summary, it generates a verifiable test case. Although the module also provides feedback on the current patch and development state later on, its main goal at this stage is to produce a consistently failing test that confirms the presence of the bug. Once the test is generated, it is executed, and the results are analyzed. If the test passes, indicating that the bug is resolved and the patch does not negatively impact the project, the current patch is finalized, and the process stops. If the test fails, the system evaluates whether the failure is due to an invalid test or an unresolved bug. Based on this, it either refines the test or produces a diagnostic report with logs and suggestions for further debugging. Throughout the process, the system also considers the current patch state to assess how each change affects the bug and the overall functionality, guiding whether to continue with the current patching method or to find a new solution.

To support automated repair, the \texttt{Patching} module implements the \textbf{Iterative Code Search and Repair} (ICSR) process. This multistep framework searches for candidate fixes and applies them iteratively. Crucially, ICSR supports rollback capabilities, allowing the system to revisit earlier stages and explore alternative strategies when the current direction is found to be ineffective or suboptimal.

\subsection{Iterative Code Search and Repair Process (ICSR)}
\label{ssec:tcam_process}

\begin{figure*}[htpb]
    \centering
    \includegraphics[width=0.83\textwidth]{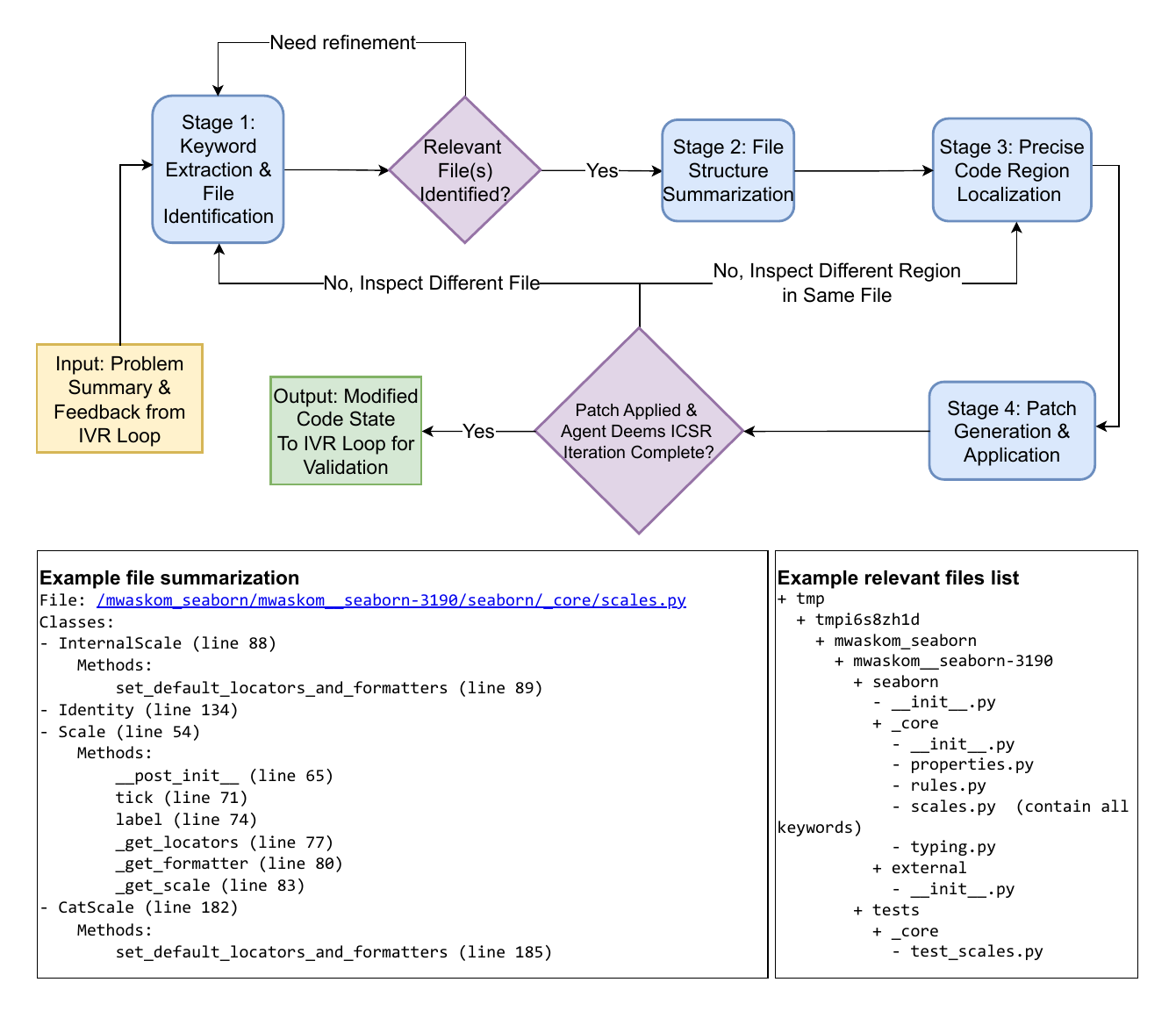}
    \caption{Iterative Code Search and Repair workflow, which includes many steps from locating to fixing errors, and example display of file summarization and relevant files list in the work.}
    \label{fig:icsr-flow}
\end{figure*}

The patching module is designed to methodically identify and apply minimal modifications necessary to resolve a reported issue. This process is organized into four sequential stages: identifying relevant files, summarizing their structure, locating precise code regions, and generating a patch. Initially, the agent extracts a list of keywords based on both the problem description and the project's directory structure. These keywords are then used to query the project tree for potentially relevant files. The resulting matches are displayed in a tree-like format as shown in Figure~\ref{fig:icsr-flow}, allowing the agent to visualize the layout of the codebase. Furthermore, the agent can reassess its choices and backtrack to refine the keyword list, leveraging feedback from previous attempts to improve the relevance of the search.

Upon identifying the appropriate files, the agent proceeds to summarize the structure of each file by listing its classes, functions, and their respective line spans. This structural overview helps the agent reason about file content without parsing the full code in one pass. Using this high-level map, the agent then inspects the internal content of specific classes and functions, enabling the precise localization of potential bug sites. If the existing implementation appears correct or irrelevant to the issue, the agent can shift its focus to alternative files or code regions using the system-provided tools. Importantly, edits are made conservatively: only one code region is modified per iteration, and viewing a different file resets all prior modifications to ensure patches remain minimal.

The agent follows a strictly linear process, advancing to the next stage only after completing the current one. However, many stages include a rollback mechanism to handle flawed decisions, such as incorrect keyword selection or file relevance. In such cases, the agent explains the mismatch between expectations and outcomes and justifies revisiting an earlier stage with improved criteria. This adaptive correction improves the robustness of the patching strategy without disrupting the structured workflow. Furthermore, the entire process is guided by React-style prompting, where reasoning and actions are tightly coupled at each step. A persistent problem summary is included in all prompts to maintain focus, and a truncation mechanism retains only the most recent exchanges to prevent history overload. When rolling back, the conversation history is trimmed to that point, removing later interactions. Reflective feedback is essential for explaining failures and planning improvements, avoiding repeated mistakes, and making informed decisions.
\section{Experiments}
\label{sec:experiments}


\subsection{Experimental Result}
\label{ssec:exp_setup}
\begin{table}[htbp]
    \centering
    \begin{tabular}{l|c}
        \textbf{Method} & \textbf{Score (\%)} \\\hline
        Moatless Tools + DeepSeekV3 & 0.00 \\
        Swe-Fixer & 24.67 \\
        \textbf{Repeton + DeepSeekR1 (Ours)} & \textbf{11.67} \\\hline
    \end{tabular}
    \caption{Performance comparison on the Swebench-lite benchmark. The score reflects the percentage of bugs successfully fixed.}
    \label{tab:experiment_results}
\end{table}

We evaluated our pipeline on the \textbf{Swebench-lite} benchmark and compared it with other open-weight coding agents. Swebench-lite includes 300 GitHub repositories with real-world software bugs and is widely used to assess automated software engineering systems. The target is to generate code patches that fix the reported issues. We compare our method against two open-weight baselines: \texttt{Moatless Tools + DeepSeekV3} and \texttt{Swe-Fixer}. \texttt{Swe-Fixer} is trained and fine-tuned specifically for Swebench-style bug fixing and uses an additional text retrieval module to enhance patch quality, though this adds complexity and reduces efficiency. Our method, in contrast, uses only general-purpose, open-weight models such as DeepSeek, without task-specific tuning or retrieval components.

As shown in Table~\ref{tab:experiment_results}, our system achieves a success rate of 11.67\%, outperforming \texttt{Moatless Tools + DeepSeekV3}, which failed to resolve any of the benchmark tasks. Although \texttt{Swe-Fixer} achieves a higher overall success rate, our pipeline demonstrates competitive performance, particularly considering its minimal architectural assumptions and the lack of specialized fine-tuning.

\section{Failure Case Analysis}
\label{ssec:failed_analysis}

\begin{table}[htpb]
    \centering
    \begin{tabular}{c|cccc}
         & Resolved & Unresolved & Empty Patch & Total \\\hline
         \# Instances & 35 & 113 & 152 & 300 \\\hline
    \end{tabular}
    \caption{Summary of model outcomes. Many failures are due to the agent not generating any patch content.}
    \label{tab:failed}
\end{table}

We analyze the failure cases of our method using DeepSeek-R1, as summarized in Table~\ref{tab:failed}. One major issue is that the model sometimes does not generate any patch. This often happens when the agent produces an overly long reasoning sequence, which causes it to run out of context and stop before producing a final answer. This highlights a risk when using general-purpose reasoning models such as DeepSeek-R1 in this type of task.

Another common failure is the agent’s inability to identify the correct files. This results in repeated unsuccessful searches without any code modifications, revealing a weakness in the pipeline’s ability to guide the agent effectively. In some cases, the agent also fails to reproduce the reported error. Since the pipeline relies on successful error reproduction to validate patches, this can cause multiple valid patches to be rejected. Eventually, only the last patch is accepted by default, even though earlier patches are often more correct. This behavior can significantly reduce overall performance.

\section{Conclusion}

We present Repeton, a fully open-source framework for autonomous software debugging. Unlike monolithic LLM-based systems, Repeton structures the debugging process as a modular patch-and-test pipeline, enabling precise, interpretable, and verifiable code edits in real-world coding projects or repositories. By combining open-source LLM reasoning with symbolic heuristics and developer tools, Repeton efficiently navigates and modifies codebases. This research also examines cases where the pipeline failed to resolve issues, providing a valuable resource for future development in this field. Finally, Repeton offers a practical and extensible foundation for LLM-driven software agents, prioritizing transparency and real-world applicability.

\bibliographystyle{ACM-Reference-Format}
\bibliography{sample-base}

\end{document}